\documentclass[desactivate]{aa}

\usepackage{newtxtext,newtxmath}
\usepackage{tablefootnote}
\usepackage[T1]{fontenc}
\usepackage[colorlinks=true,linkcolor=blue,citecolor=blue,urlcolor=blue]{hyperref}

\DeclareRobustCommand{\VAN}[3]{#2}
\let\VANthebibliography\thebibliography
\def\thebibliography{\DeclareRobustCommand{\VAN}[3]{##3}\VANthebibliography}

\usepackage{graphicx}	
\usepackage{amsmath}	

\defcitealias{fonte23}{Paper I}

\begin{document}

\title{The effect of vertical mixing on oxygen sequestration by refractory elements in the atmosphere of giant planets}

\author{Paolo Simonetti\inst{1,2,3}
\and Diego Turrini\inst{2,4}
\and Eugenio Schisano\inst{2,5}
\and Stavro L. Ivanovski\inst{1,3}
\and Romolo Politi\inst{2,5}
\and Sergio Fonte\inst{5}
\and Elenia Pacetti\inst{5}
\and Sergio Molinari\inst{5}
\and Matteo Brogi\inst{6,4}
\and Danae Polychroni\inst{1,2,4}
\and Michele Zusi\inst{5}
\and Valeria Cottini\inst{7}}

\institute{INAF - Osservatorio Astronomico di Trieste, Via G. B. Tiepolo 11, 34143 Trieste, Italy
\and
National Research Centre for High Performance Computing, Big Data and Quantum Computing (ICSC), Via Magnanelli 2, 40033 Casalecchio di Reno (BO), Italy
\and
Institute for the Fundamental Physics of the Universe (IFPU), Via Beirut 2, 34014 Trieste, Italy
\and
INAF - Osservatorio Astrofisico di Torino, Via Osservatorio 20, 10025 Pino Torinese (TO), Italy
\and
INAF - Istituto di Astrofisica e Planetologia Spaziali, Via del Fosso del Cavaliere 100, 00133 Rome, Italy
\and
Universita' di Torino, Dept. of Physics, Via P. Giuria 1, 10125 Turin, Italy
\and
Agenzia Spaziale Italiana, Via del Politecnico snc, 00133 Rome, Italy
}

\date{Accepted XXX. Received YYY; in original form ZZZ}

\abstract{The most widely used route to study the formation history of giant exoplanets is the C/O ratio, usually derived from H$_2$O, CO, CO$_2$ and CH$_4$. However, refractory elements can sequester a significant fraction of O, causing the observed C/O ratio to be overestimated with respect to the intrinsic one. Previous calculations of this bias relied on chemical equilibrium models. Here we quantify the impact of vertical mixing using the 1-D chemical kinetics code VULCAN, expanded with a Mg-Si-Fe network including the condensation and settling of enstatite, forsterite and metallic iron. We compute 168 models spanning equilibrium temperatures $T_{eq} \in$  [700, 1500] K, eddy diffusion coefficients $K_{zz} \in$ [$10^6$, $10^{10}$] cm$^2$ s$^{-1}$ and metallicities $Z \in$ [0.9, 7.2] $Z_\odot$, derived self-consistently from planet formation simulations. We find that vertical transport causes magnesium silicates to \textcolor{black}{destabilize} already at $T_{eq} \sim 900$ K regardless of metallicity, releasing oxygen to H$_2$O and to minor gas-phase carriers such as SiO and Mg(OH)$_2$. Therefore, while equilibrium models predict a nearly constant C/O \textcolor{black}{overestimation} up to 1100–1300 K, in disequilibrium it drops sharply between 900 and 1200 K. In this transitional range the C/O \textcolor{black}{overestimation} is controlled by $K_{zz}$ and is partially degenerate with metallicity, while outside it is robust against both mixing intensity and the probed atmospheric layer, ranging from 14.5-23\% at 700 K to 4.5-7\% at 1500 K. Metallic iron condensation prevents Fe from sequestering oxygen, lowering the \textcolor{black}{overestimation} by about one third with respect to previous estimates. Finally, CH$_4$, CO$_2$ and the CH$_4$/CO ratio are most sensitive to $K_{zz}$ in the transitional range, and can provide a constraint on vertical mixing.}

\keywords{planets and satellites: atmospheres -- planets and satellites: composition -- planets and satellites: formation -- astrochemistry}

\titlerunning{Oxygen deficit in vertically mixed atmospheres}
\authorrunning{Simonetti et al.}

\maketitle



\section{Introduction}

Thanks to the rapidly increasing number and quality of available exoplanetary spectra a transition from the detections of species to the quantitative compositional characterization of known objects is well underway. While early investigations were able to identify only especially opaque \citep[e.g. Na,][]{charbonneau02} or abundant \citep[e.g. H$_2$O,][]{tinetti07} chemical species in exoplanetary atmospheres, the continuous improvements in both spectral resolution and sensitivity are now allowing to retrieve the concentrations of several molecules at once \citep{cabot19,giacobbe21,guilluy22,rustamkulov23,alderson23,yang24}, even in cooler and smaller targets \citep{madhusudhan23,beatty24}. The recently announced discovery of silicate clouds in the atmosphere of the exoplanet YSES-1 c \citep{hoch25} further expanded observations to the domain of condensates. These advances make possible to characterize the elemental composition of a larger and larger number of Jupiter-like objects with an unprecedented degree of precision.

The existence of the so-called Hot Jupiters in close orbits around their parent star and the difficulty in explaining how they could have formed in situ \citep{rafikov05,schlichting14,lee16} strongly support the idea that planets can migrate inside a planetary system. Migration can happen at several stages in a planet's history, such as during the formation via interactions with the protoplanetary disc \citep{goldreich80,nelson00,alexander09,baruteau14}, or at a later time because of planet-planet interactions \citep{turrini20,turrini22,zhu21,hamer22}. While alternatives to this picture do exist \citep{batygin16,boley16,poon21}, migration is generally considered as the best solution to the existence of short period massive planets.

One way to shed light on the many uncertainties related to the formation and migration processes \citep[see e.g. the reviews by][]{dawson18,fortney21} is to reconstruct the evolutionary history of known exoplanets by using their compositions \citep{oberg11,mordasini16,madhusudhan19,oberg21,turrini21,turrini22,knierim22,pacetti22,pacetti25,pacetti26}. This is possible since protoplanetary discs exhibit chemical gradients, generally described in terms of snow lines, i.e. radial boundaries identifying the orbital regions where the disc temperature gradient causes the condensation of chemical species such as H$_2$O, CO$_2$, NH$_3$ and CO in grains. These grains aggregate to form planetary embryos which will then grow by accretion of nearby dust and planetesimals. Once their mass reaches a certain threshold value, these objects will undergo a runaway accretion of the surrounding gas, producing Saturn- and Jupiter-like objects \citep{johansen17,tanaka20}. Specific elemental ratios are thus imprinted on the atmospheres of these planets, that can then be observed and interpreted \citep[e.g.][]{moses13,line14,pinhas19,spake21,turrini21,pacetti22,pacetti25,pacetti26}. The most widely investigated in this sense is the C/O ratio, hereupon called $R_{\text{obs}}$, since C and O are the most abundant heavy elements and thus are the easiest to observe with the current instrumentation \citep[see e.g. the review by][]{madhusudhan19}. The value of $R_{\text{obs}}$ is derived by estimating the volume mixing ratios (VMR) of H$_2$O, CO$_2$, CO and CH$_4$ by the strength of their near- and mid-infrared absorption bands. However, O is a very reactive element that can be readily sequestered by other, more difficult to observe species. Among them, there are the oxides and hydroxides of relatively abundant refractory elements such as Si, Mg and Fe, which can be present both in gas phase and condensed in grains. At temperatures below 1100-1200 K, around 22\% of O atoms in a solar-composition atmosphere can be bound to refractories, thus skewing $R_{\text{obs}}$ \citep{burrows99,fegley10,visscher10}. This bias, known as oxygen deficit, has been investigated in the case of warm and hot Jupiters using a chemical equilibrium code for a variety of formation and migration histories by \citet{fonte23} \citepalias[hereafter,][]{fonte23} who concluded that, due to the non-solar composition of these planets' atmospheres, it can reach values around 30-40\% at temperatures below 1000 K. The same study identified a new class of transitional hot (900-1100 K) exoplanets, where oxygen deficit impacts the transition between CH$_4$- to CO-dominated atmospheres.

Chemical disequilibrium is exoplanetary atmospheres is expected to play an important role in shaping their chemical makeup \citep{spiegel09,moses11,visscher11,venot12,line13,drummond16,molaverdikhani19}. Examples of observations for which it has been invoked are the presence of HCN and NH$_3$ in Hot Jupiters \citep{macdonald17} and the CH$_4$ depletion in Warm Neptunes \citep{kreidberg18}. A variety of processes can drive this intrinsically 3D phenomenon, including dayside-nightside temperature differences, vertical mixing, highly energetic radiation and atmospheric escape. Models with varying levels of complexity have been deployed to explore these effects, from fully integrated global circulation, chemical kinetics and radiative transfer codes \citep[e.g.][]{drummond20,zamyatina23} to partially parameterized or semi-analytical approaches that focus on a specific aspect, such as the atmospheric transport \citep{mendonca18,beltz21,baeyens21}, photochemistry \citep{tsai24} or photoevaporation \citep{rumenskikh22}. Both self-consistent but computationally expensive coupled models and much faster single-column calculations that can be used to cover the planetary parameter space are required to disentangle this complex problem, and to interpret the increasing volume of high-quality observational data on exoplanetary atmospheres.

In this work, we make use of the chemical kinetics code Vulcan \citep{tsai17,tsai21} complemented with an expanded chemical reactions network and updated physics recipes for gravitational settling \citep{simonetti26}. This expanded version of the code allows for a chemistry that includes Mg, Si, Fe and their main carrier species. In particular, we focus on studying how vertical mixing-driven chemical disequilibrium affects the oxygen deficit and the VMRs of the observationally relevant molecules CO$_2$, CH$_4$ in short-period Jupiter-like planets. As for the initial elemental abundances, we consider the same set of compositions as in \citetalias{fonte23}, which have been derived self-consistently by the planetary formation and migration model presented in \citet{turrini21} and \citet{pacetti22} adopting a Solar-like metallicity and composition for the parent star.
We focus on the atmospheric layers that are currently observed by the NASA/ESA/CSA mission JWST \citep{greene16} and that will be probed by the ESA mission Ariel \citep{tinetti18}, i.e. between 0.1 and 1 mbar for transmission spectroscopy, and between 0.1 and 1 bar for emission spectroscopy.

The paper is organized as follows. In Section \ref{sec:model} we briefly describe the chemical kinetics model, the chemical network and the adopted atmospheric profiles and compositions. In Section \ref{sec:results} we show our results, both in terms of major O-bearing species and other molecules of interest. In Section \ref{sec:discussion} we discuss the implications of vertical mixing on the retrieved $R_\text{obs}$ values and in Section \ref{sec:conclusions} we summarize the main conclusions of this work.

\section{Model}
\label{sec:model}

We investigate the vertical mixing ratio of the main O-bearing species using Vulcan \citep{tsai17,tsai21}. Vulcan is an open-source photochemical kinetics code with vertical transport that calculates the steady state composition of an atmosphere by solving a chemical reactions network by means of the Rosenbrock method. In contrast with chemical equilibrium solvers such as FastChem \citep{stock18,stock22,kitzmann24} or GGChem \citep{woitke18}, Vulcan results include the effects of chemical disequilibria caused by vertical mixing, photodissociation and particle settling of the condensed phase. By specifying appropriate boundary conditions, Vulcan can also capture the effect of atmospheric escape and the interaction between the atmosphere and the planetary interiors.

Specifically, the code works by integrating a set of mass continuity equations, one for each of the tracked chemical species $X$. Production and destruction rates of $X$ are calculated via the generalized Arrhenius equation \citep{heng16} of the reactions involving $X$, which in general are derived experimentally, while the vertical transport term depends on molecular diffusion, eddy diffusion and vertical advection. Condensation and evaporation are treated as reactions involving the gas phase and the condensed phase of the same species whose rate is determined following the formulation presented in \cite{seinfeld16}. Particle settling also contributes indirectly to the flux term by moving condensates downward. As such, the concentration of a given species at a given height in the atmosphere is essentially the outcome of the interplay between the timescales associated to production and destruction, which depend on the magnitude of the forward and backward reaction rates, and the timescale associated to the transport processes.

In \citet{simonetti26}, we modified Vulcan both in terms of chemical network and physical model. In particular, we expanded the list of species to include 18 oxides and hydroxides of Mg, Si and Fe, adapting it to treat the condensation and dissolution of the two main magnesium silicates whose formation is expected in substellar atmospheres (enstatite and forsterite). In terms of physics, we refined the treatment of the gravitational settling of condensate particles. Initialization of molecular VMRs is now done by running a modified version of FastChem 3 \citep{kitzmann24} including only the species that are also treated in Vulcan. All equilibrium results shown in this paper have been obtained by this version of FastChem. We direct the reader to \citet{simonetti26} for the implementation details and the validation of this framework.

\subsection{Vertical mixing}
\label{ss:vertical_mixing}

In Vulcan, vertical mixing is modeled via three terms: advection, eddy diffusion and molecular diffusion. These terms capture, respectively, the bulk gas motion in convective cells, the small-scale turbulent fluid mixing and the diffusion by molecular-level random walk at low pressures. At the atmospheric depth investigated in this work vertical mixing is dominated by the first two terms and we will focus specifically on the second one.

The eddy diffusion is an idealization representing atmospheric vertical mixing in a simplified and computer-friendly way adopted by most 1D chemical kinetics models \citep[see e.g.][and references therein]{tsai17}. Despite its widespread usage, the link between the value of the eddy diffusion coefficient $K_{zz}$ and the underlying physics of the atmosphere is not robustly defined. Rather than deriving it from first principles, the value of the eddy diffusion coefficient $K_{zz}$ is chosen \textit{a posteriori} to match the vertical VMR profile of some chemical tracer \citep{prinn77} or the average vertical diffusive mass flux \citep{chamberlain87} as derived either from observations of the atmosphere of Solar System bodies or calculated by General Circulation Models (GCMs). Both methods -chemical tracers and average diffusive mass flux- have been shown to be generally in agreement, even though tracers with different chemical timescales cause the resultant $K_{zz}$ to spread over up to one order of magnitude \citep{visscher11,parmentier13,komacek19}. Observations of the atmospheres of Venus, Earth and Jupiter allowed to derive $K_{zz}$ values in the [$4 \times 10^4$, $4 \times 10^6$] cm$^{2}$ s$^{-1}$ \citep{zhang12}, [$5 \times 10^{3}$, 10$^{5}$] cm$^{2}$ s$^{-1}$ \citep{massie81} and [10$^{3}$, 10$^{7}$] cm$^{2}$ s$^{-1}$ ranges \citep{moses05}, respectively. The highest values are associated with the troposphere and the mesosphere/thermosphere, while the lowest ones are found in the stratosphere. Warm and hot Jupiter-like cases have been investigated via GCM simulations, which yielded globally averaged vertical diffusivities in the [10$^{8}$, $10^{10}$] cm$^{2}$ s$^{-1}$ interval for tidally-locked planets in the 500-1500 K equilibrium temperature range \citep{komacek19}. Starting from previous models referred specifically to HD 189733 b and HD 209458 b by \citet{showman09}, \citet{moses11} derived a $K_{zz}$ in the [10$^{8}$, $2 \times 10^{10}$] and [$2 \times 10^{6}$, $3 \times 10^{11}$] cm$^{2}$ s$^{-1}$ ranges, respectively. Using a different GCM diagnostic, \citet{heng11} derived an average $K_{zz} \sim 10^{10}$ cm$^{2}$ s$^{-1}$ for HD 209458 b, while calculations based on (simulated) tracer concentrations by \citet{parmentier13} applied to the same planet found a pressure-dependent mixing that varies between $5 \times 10^8$ cm$^{2}$ s$^{-1}$ at 1 bar and $2 \times 10^{11}$ cm$^{2}$ s$^{-1}$ at 10$\mu$bar, which was later confirmed by \citet{zhang18} for the [0.001, 1] bar region. Following these results, we decided to test three $K_{zz}$ values in this work: $10^6$, $10^8$ and $10^{10}$ cm$^{2}$ s$^{-1}$, kept constant at all pressures.

\subsection{Chemical network}
\label{ss:network}

For this study, we adopt the \texttt{SILANE} network developed in \citet{simonetti26}. This network is tailored to study how Mg, Si and Fe, which are the most cosmically abundant refractories, interact with the stock of volatile species and explore how O is partitioned among its main carriers. \texttt{SILANE} is constituted of two sub-networks. The first one is the $\texttt{SNCHO\_photo\_network\_2024}$ file distributed with the Vulcan code itself and already tested against other photochemical kinetics codes in \cite{tsai21}. It includes 437 second-order reactions, 77 third-order reactions and 59 photodissociation reactions for 88 elements and molecules: H, O, C, N, S, H$_2$, OH, H$_2$O, HO$_2$, H$_2$O$_2$, CH, CH$_2$, CH$_3$, CH$_4$, C$_2$, C$_2$H$_2$, C$_2$H, C$_2$H$_3$, C$_2$H$_4$, C$_2$H$_5$, C$_2$H$_6$, CO, CO$_2$, CH$_2$OH, H$_2$CO, HCO, CH$_3$O, CH$_3$OH, CH$_3$CO, O$_2$, H$_2$CCO, HCCO, NH, CN, HCN, NO, NH$_2$, N$_2$, NH$_3$, N$_2$H$_2$, N$_2$H, N$_2$H$_3$, N$_2$H$_4$, HNO, H$_2$CN, HNCO, NO$_2$, N$_2$O, C$_4$H$_2$, CH$_2$NH$_2$, CH$_2$NH, CH$_3$NH$_2$, CH$_3$CHO, HNO$_2$, NCO, HC$_3$CN, CH$_3$CN, CH$_2$CN, C$_2$H$_3$CN, SH, HSO, H$_2$S, C$_3$H$_3$, C$_3$H$_2$, C$_3$H$_4$, C$_6$H$_5$, C$_6$H$_6$, S$_2$, S$_3$, S$_4$, S$_8$, SO, CS, COS, CS$_2$, NS, HS$_2$, SO$_2$, HCS, C$_4$H$_3$, C$_4$H$_5$, S$_2$O, CH$_3$SH, CH$_3$S, $^1$O, $^1$CH$_2$, N($^2$D), plus the non-reactive He. The second one is described in Appendix B of \citet{simonetti26} and includes 46 first, second and third-order reactions and 18 new species: Mg, Si, Fe, MgO, MgOH, Mg(OH)$_2$, FeO, FeO$_2$, Fe(OH)$_2$, SiS, SiO, SiO$_2$, SiH, SiH$_2$, SiH$_3$, SiH$_4$, MgSiO$_3$ and Mg$_2$SiO$_4$. Fe can condense following the pressure-temperature relation presented in \cite{visscher10}. MgSiO$_3$ and Mg$_2$SiO$_4$ exist both in the condensate form and in a virtual, gaseous form that rapidly dissociates in simpler Mg and Si species when the respective evaporation conditions are met. The MgSiO$_3$ and Mg$_2$SiO$_4$ condensation curves have been taken from \cite{visscher10}.

\subsection{Grain settling}
\label{ss:settling}

Grain radius is not self-consistently calculated and dynamically evolved in Vulcan. Rather, the user specifies a single value for each condensate that will be used to calculate the evaporation rate and the settling velocity. In this work, as in \citet{simonetti26}, we adopt a common radius for all condensate particles of 1 $\mu$m, which is the order of magnitude of the maximum size of grains in the outer regions of substellar atmospheres \citep{woitke03}. Evaporation is treated following Eq.~14 of \citet{tsai21}, which is itself taken from \citet{seinfeld16}, and describes the growth and evaporation rates of a particle in the continuum regime. Condensation nuclei are assumed to be always present.

For the settling velocity we adopt the approach described in \citet{simonetti26}, which is itself taken from \citet{huang24}. The settling velocity is calculated as the harmonic mean between the Epstein and the Stokes settling velocities, which gives the model the ability to satisfactorily model sedimentation both in the free molecular flow, and the in continuum fluid regime.

\subsection{Atmospheric pressure-temperature profiles and planetary parameters}
\label{ss:atmo_profiles}

\begin{figure}
\includegraphics[width=0.45\textwidth]{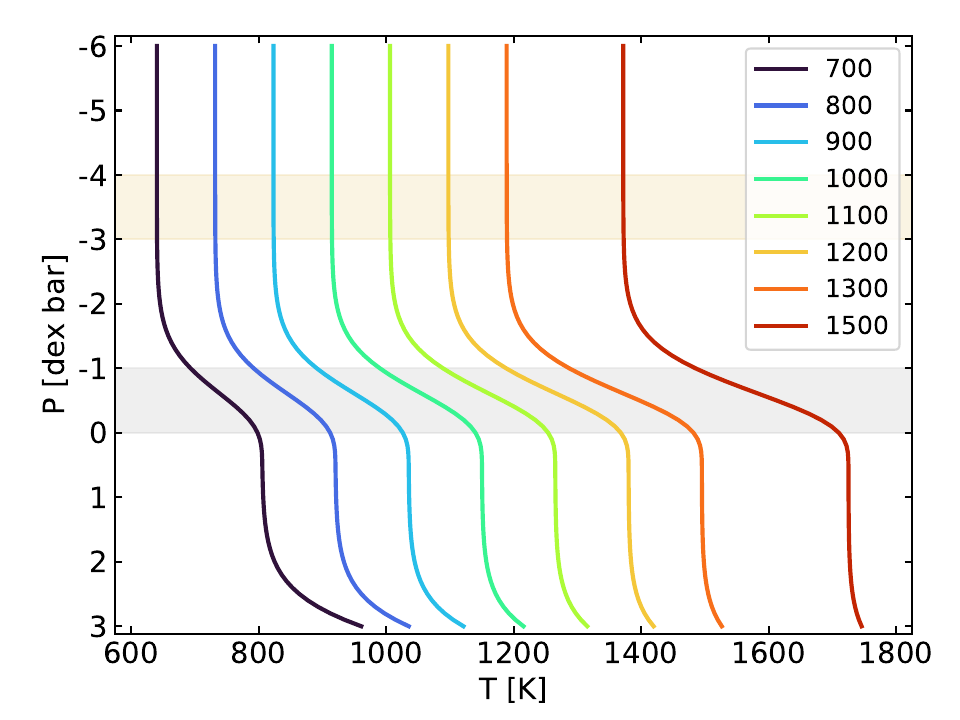}
\caption{Vertical temperature profiles used in this study. The legend reports the planetary equilibrium temperatures associated with each profile. The gold and silver shaded areas underline the atmospheric layers between 0.1 mbar and 1 mbar, and between 0.1 and 1 bar, which corresponds, respectively, to the regions of maximum sensitivity for transit and emission spectroscopy, for both the ESA Ariel \citet{tinetti18} mission and the NASA/ESA/CSA JWST \citet{greene16} mission.}
\label{fig:temp_profiles}
\end{figure}

As in \citetalias{fonte23}, in this work we adopt the pressure-temperature relationship between the temperature $T_\text{p}$ at a given vertical optical depth $\tau$, for a specified internal temperature $T_{\text{int}}$, equilibrium temperature $T_{eq}$ and visible-to-thermal infrared opacity ratio $\gamma=\kappa_{\text{vis}}/\kappa_{\text{th}}$ derived in \citet{guillot10}:
\begin{equation}
\label{eq:guillot}
\overline{T_\text{p}^4} = \frac{3}{4}T_{\text{int}}^4 \biggl(\frac{2}{3}+\tau\biggl) + \frac{3}{4}T_{eq}^4 \biggl(\frac{2}{3}+a_1(\tau,\gamma)+a_2(\tau,\gamma)\biggl)
\end{equation}
The $a_1$ and $a_2$ functions parameterize the effects of opacities and horizontal redistribution on the temperature at a specific $\tau$ level of the atmosphere. This profile should be intended as a global average for the planet, hence the use of the bar sign over $T_\text{p}$. Under the hypothesis of hydrostatic equilibrium and assuming a plane-parallel geometry, the pressure $P$ and $\tau$ are related at any given depth by:
\begin{equation}
\tau = \frac{\kappa_{\text{th}}P}{g}
\end{equation}
where $\kappa_{\text{th}}$ is the opacity in the thermal infrared and $g$ is the gravity acceleration, supposed constant throughout the atmosphere. We assume $T_{\text{int}}=100$ K and $\kappa_{\text{th}}=0.01$ cm$^2$g$^{-1}$. In order to calculate $a_1$ and $a_2$, we adopt an opacity in the visible band $\kappa_{\text{vis}}=0.004$ cm$^2$g$^{-1}$. \citet{guillot10} showed that these choices are appropriate to reproduce the pressure-temperature profile of HD 209458 b found by detailed GCM models for a Solar-composition atmosphere. Since this parameterization is widely adopted by retrieval codes \citep[e.g.][]{molliere19,ormel19,zhang20}, our results can be readily applied to observational studies in literature. Oxygen depletion has been shown to be strongly dependent on $T$ \citepalias{fonte23}. As such, we test eight different atmospheric vertical profiles, one for each value of the equilibrium temperature $T_{eq}$ in the set $[700,800,900,1000,1100,1200,1300,1500]$ K, which are shown in Fig.~\ref{fig:temp_profiles}. The $P$ at the top and at the bottom of the atmospheric column are, respectively, 1 $\mu$bar and 1000 bar and each profile contains 91 pressure points (10 per order of magnitude).

In principle, $\kappa_{\text{th}}$ and $\kappa_{\text{vis}}$ change in response to variations of the chemical composition, which in turn depends on the chosen metallicity $Z$ and $T_{\text{eq}}$. For example, \cite{drummond18} showed that raising the metallicity from Solar to ten times Solar in the warm sub-Neptune atmosphere of GJ 1214 b increases the temperature at the 100 bar level from $\sim$1170 K to $\sim$1390 K. Infrared and visible opacity change also along the atmospheric column, in response to the specific chemical composition at each layer. Exploring in a fully self consistent manner these effects requires the use of a coupled photochemical-radiative transfer model, which falls outside the scope of our present modeling effort. However, we test the impact of doubling $\kappa_{\text{th}}$, which increases the temperature in the 100-1000 bar region by 100-200 K, depending on $T_{eq},$ for the subset of cases with the highest metallicity. The results of this test are discussed in Sect.~\ref{ss:metopac}.

Concerning the mass and radius, we follow \citetalias{fonte23} and model our Warm/Hot Jupiter planet after HAT-P-5 b, as this planet is characterized by physical parameters compatible with those of the simulated planets from \citet{turrini21} and \citet{pacetti22}. As such we adopted mass and radius values respectively equal to 0.98 M$_{\text{J}}$ and 1.21 R$_{\text{J}}$ \citep{stassun17}\footnote{\url{https://exoplanetarchive.ipac.caltech.edu/overview/HAT-P-5b}}. The resultant gravity acceleration is $1734$ cm s$^{-2}$.

\subsection{Planetary elemental compositions}
\label{ss:compositions}

\begin{table}
\centering
\caption{Elemental abundances in number used to initialize Vulcan, expressed in $\log_{10}(X/H)+12$.}
\begin{tabular}{lcccccc}
\hline
Element & \multicolumn{6}{c}{Initial distance $d_0$ of the planetary embryo (AU)} \\
 & 5 & 12 & 19 & 50 & 100 & 130 \\
\hline
He & 10.98 & 10.98 & 10.98 & 10.98 & 10.98 & 10.98 \\
C & 8.51 & 8.66 & 8.73 & 8.96 & 9.25 & 9.35 \\
N & 7.86 & 7.90 & 7.98 & 8.13 & 8.34 & 8.41 \\
O & 8.69 & 8.89 & 8.98 & 9.25 & 9.55 & 9.64 \\
Mg & 7.43 & 7.79 & 7.89 & 8.17 & 8.48 & 8.57 \\
Si & 7.35 & 7.71 & 7.81 & 8.09 & 8.40 & 8.49 \\
S & 6.96 & 7.32 & 7.42 & 7.70 & 8.01 & 8.10 \\
Fe & 7.31 & 7.67 & 7.77 & 8.05 & 8.36 & 8.45 \\
\hline
Z/Z$_\odot$ & 0.9 & 1.4 & 1.7 & 3.0 & 5.8 & 7.2 \\
O/O$_\odot$ & 1.01 & 1.58 & 1.97 & 3.63 & 7.20 & 8.87  \\
C/C$_\odot$ & 1.19 & 1.69 & 1.98 & 3.42 & 6.67 & 8.27 \\
Fe/Fe$_\odot$ & 0.65 & 1.46 & 1.86 & 3.56 & 7.18 & 8.87 \\
\hline
C/O & 0.650 & 0.587 & 0.553 & 0.518 & 0.510 & 0.512  \\
\hline
\end{tabular}
\label{tab:element_setting}
\end{table}

\begin{figure*}
\centering
\begin{minipage}[c]{0.81\textwidth}
\centering
\includegraphics[width=0.99\textwidth]{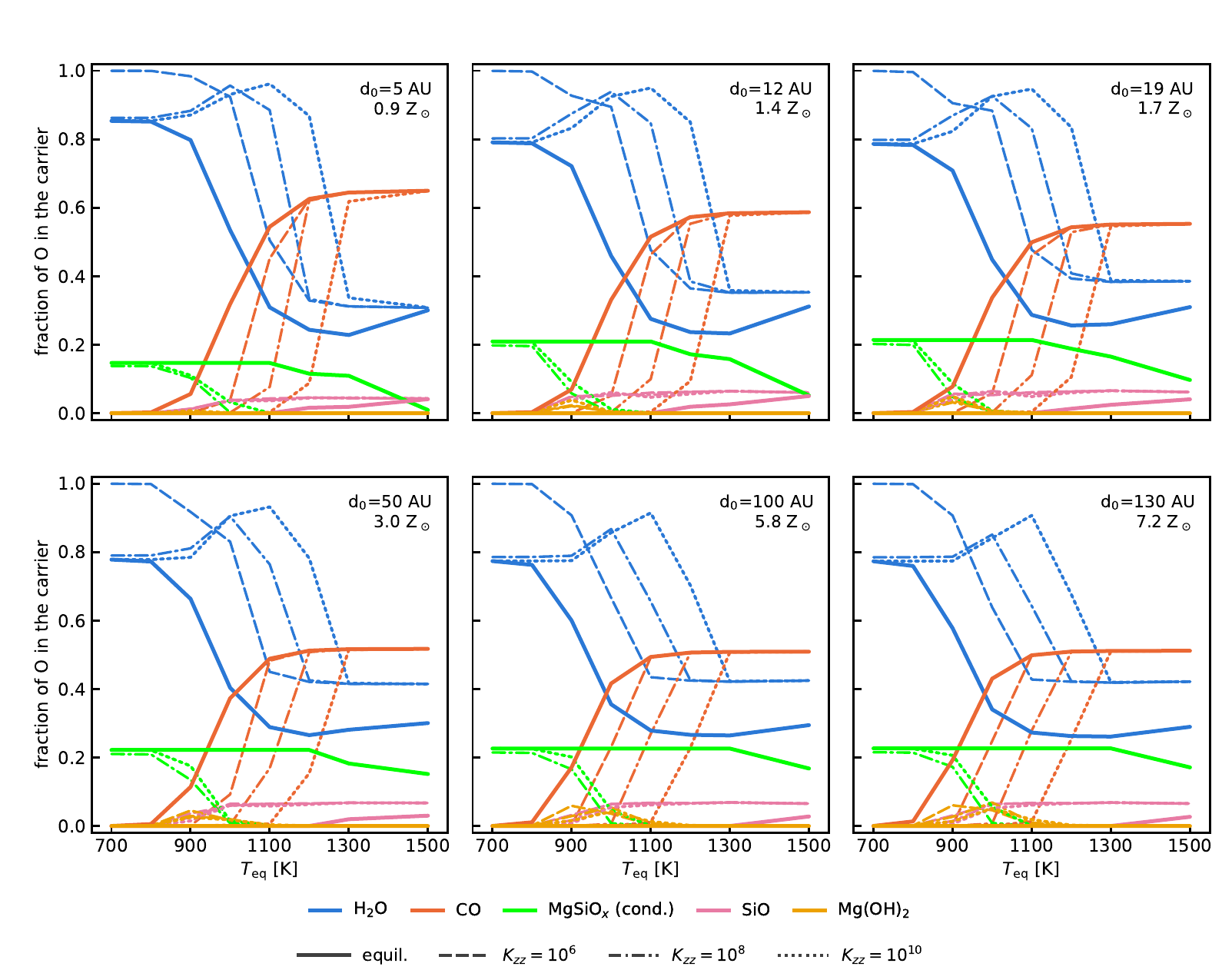}
\end{minipage}%
\hfill
\begin{minipage}[c]{0.189\textwidth}
\small
\caption{The partition of oxygen in the lower atmosphere among its major carriers as a function of equilibrium temperature for the six compositional scenarios explored here. Solid lines refer to the equilibrium chemistry results. Dashed, dash-dotted and dotted lines refer, respectively, to cases with $K_{zz}$ equal to 10$^6$, 10$^8$ and 10$^{10}$ cm$^2$s$^{-1}$. To improve readability, the oxygen fractions carried by enstatite and forsterite have been summed together (in green).\label{fig:oxypart_lower}}
\end{minipage}
\end{figure*}

The elemental abundances adopted in this work are the same as those in \citetalias{fonte23}, which have been calculated by coupling the planetary formation simulations by \cite{turrini21} with the updated compositional model of the protoplanetary disc mid-plane by \cite{pacetti22} in the scenario of chemical inheritance of the ices from the interstellar medium and low ionization of the disc midplane. The code used for the n-body simulation is the Mercury-Ar$\chi$es model described in \cite{turrini26}, which allows us to quantify the differential accretion of gas, ices, organics and refractories during the growth and migration of giant planets.

In the simulations, a Mars-sized planetary embryo (with an initial mass of 0.1 M$_\oplus$) is placed at a given initial distance $d_0$ and evolved up to a final mass of 317.8 M$_\oplus$ (i.e. a Jovian mass). The $d_0$ values tested are 5, 12, 19, 50, 100 and 130 AU, with zero initial inclination and initial eccentricity equal to 0.001 to account for the damping effects of the disc \citep{cresswell08}. The total formation time is the same for all simulations and set to 3 Myr. During this time the planet migrates inward, reaching a final semi-major axis of 0.4 AU. Due to the different starting position with respect to the multiple ice lines identified e.g. in \cite{pacetti22}, each planetary embryo accretes different amounts of volatiles and refractories, as reported in Table \ref{tab:element_setting}. The protoplanetary disc is assumed to have Solar elemental abundances as reported by \cite{asplund09} and \cite{scott15a,scott15b}. Disc material is assumed to be completely broken down to its elemental components during the accretion due to the very high temperatures reached \citep{lissauer09,dangelo21}. Elements then recombine in the atmospheres of the planets once they are sufficiently cooled.

The disc and planet formation in \cite{turrini21} and \cite{pacetti22} focus on the differential accretion of four of the most cosmically abundant heavy elements: C, N, O and S. Following \cite{turrini21} and \citetalias{fonte23}, the other refractories are quantified by multiplying the S abundance for the stellar X/S ratio, where X is the specific refractory element we are interested in. Since S is mainly carried by rocks in protoplanetary disks, it acts as a direct tracer of refractory enrichment, justifying this assumption.

\section{Results}
\label{sec:results}

\begin{table}
\centering
\caption{Model settings for the Vulcan code used in this work.}
\begin{tabular}{ll}
\hline
Parameter & Choice \\
\hline
Molecular diffusion & yes, H$_2$ as background gas \\
Thermal diffusion & yes \\
Eddy diffusion & [10$^6$, 10$^8$, 10$^{10}$] cm$^2$ s$^{-1}$ \\
Advection & 0 cm s$^{-1}$\\
Photochemistry & no \\
Condensation & Fe, MgSiO$_3$, Mg$_2$SiO$_4$ \\
Grain radius & 1 $\mu$m \\
Upper boundary & zero flux \\
Lower boundary & zero flux \\
\hline
\end{tabular}
\label{tab:model_setting}
\end{table}

The settings of the Vulcan model used in this work are summarized in Table \ref{tab:model_setting}. We considered H$_2$ to be the dominant background gas for molecular diffusion calculations, which is reasonable for the chosen $T_{eq}$ interval, i.e. [700, 1500] K. We tested three different values for the eddy diffusion coefficient $K_{zz}$: 10$^6$, 10$^8$ and 10$^{10}$ cm$^2$s$^{-1}$, which explore the expected range of values for warm and hot Jupiter atmospheres (see Sect.~\ref{ss:vertical_mixing}). $K_{zz}$ is taken constant throughout the atmosphere. \textcolor{black}{We run 144 combinations of elemental abundances, equilibrium temperatures and vertical mixing strengths, plus 24 extra cases on a modified T-P profile, for a total of 168 cases.} Condensation and sedimentation are considered for Fe, MgSiO$_3$ and Mg$_2$SiO$_4$. While they are not the only species that might condense in the explored interval range, they are expected to be the most abundant ones by chemical equilibrium calculations \citep{visscher10,fegley10}. Photochemistry is not taken into account. Since we are interested in the regions of maximum sensitivity of the ESA Ariel and the NASA/ESA/CSA JWST missions, our results refer to the vertically integrated VMRs of molecules in the [0.1, 1] bar and in the [$10^{-3}$, $10^{-4}$] bar intervals. Throughout this paper, these regions will be referred as the lower and upper atmospheric layers.

\subsection{Oxygen partition in the lower atmosphere}
\label{ss:results_lower}

\begin{figure}
\centering
\includegraphics[width=0.8\columnwidth]{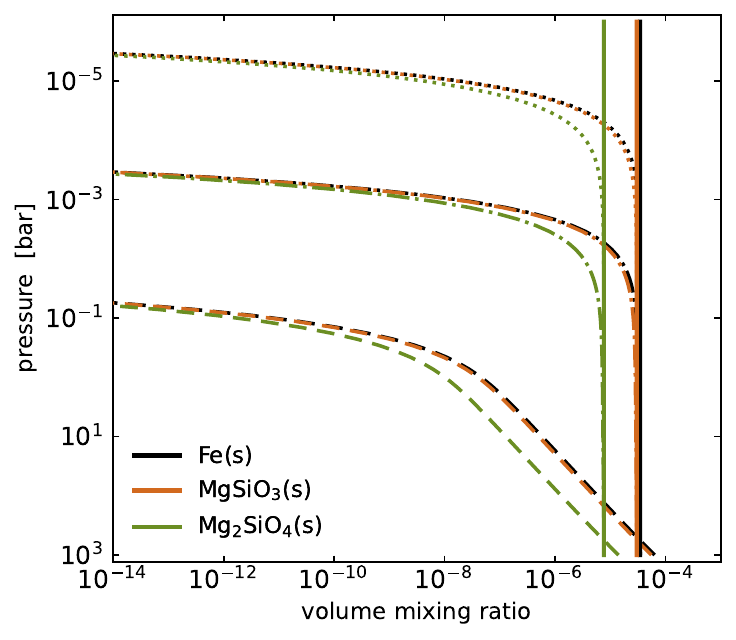}
\caption{The condensates profiles for the $T_{eq}=700$ K, $Z=0.9\,Z_\odot$ case in the equilibrium (solid) and disequilibrium models. Dashed, dash-dotted and dotted lines refer, respectively, to cases with $K_{zz}$ equal to 10$^6$, 10$^8$ and 10$^{10}$ cm$^2$s$^{-1}$.}
\label{fig:vertprof_condensates}
\end{figure}

\begin{figure}
\centering
\includegraphics[width=0.85\columnwidth]{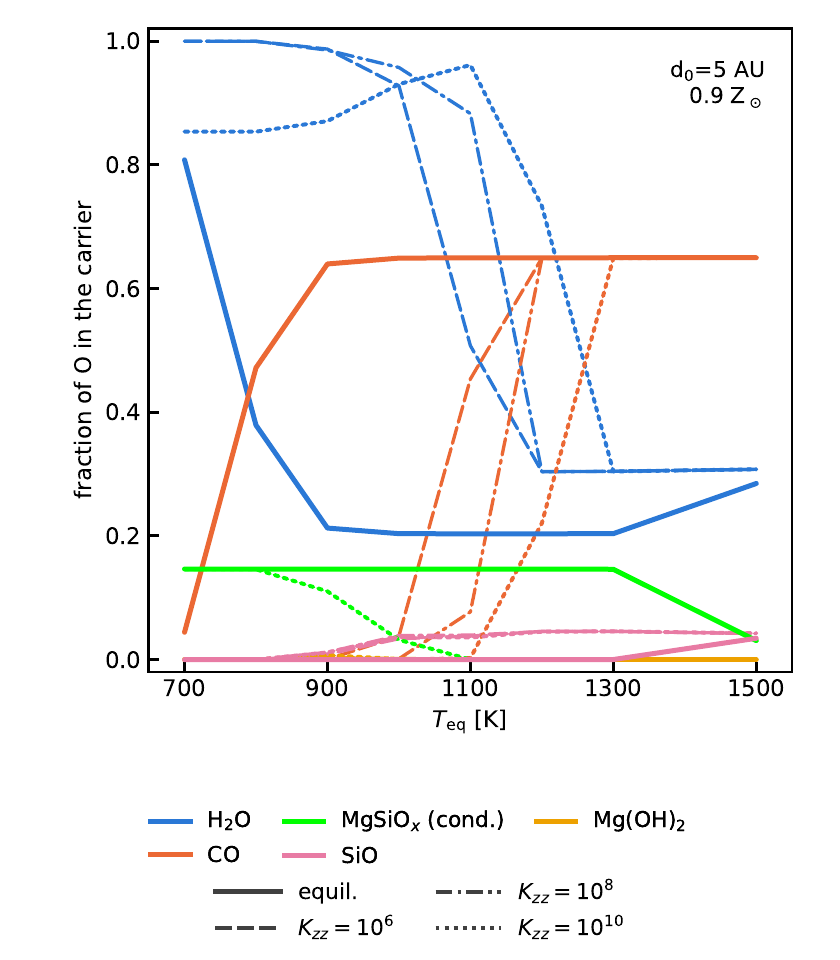}
\caption{The partition of oxygen in the upper atmosphere among its major carriers as a function of equilibrium temperature for the lowest metallicity case (i.e., the 5 AU scenario). Solid lines refer to the equilibrium chemistry results. Dashed, dash-dotted and dotted lines refer, respectively, to cases with $K_{zz}$ equal to 10$^6$, 10$^8$ and 10$^{10}$ cm$^2$s$^{-1}$. To improve readability, the oxygen fractions carried by enstatite and forsterite have been summed together (in green).}
\label{fig:oxypart_upper}
\end{figure}

The C/O ratio is both the most easily accessible and the most studied atmospheric compositional diagnostic for its links to the formation history of the planet. The goal of this paper is to explore the impact of vertical mixing on oxygen sequestration by species that are not generally observed, and that can skew the C/O ratio.
To this end, the natural starting point is to study how O is partitioned among its main carriers at the two observable layers within the atmosphere. The quantity we track is [X]$_\text{O}$, that is the fraction of O carried by species X in a given layer.

Fig.~\ref{fig:oxypart_lower} shows the O partition among its main carriers (i.e., those with [X]$_\text{O}$>0.01) in the [0.1, 1] bar atmospheric layer, over the entire [$T_{eq}$, $Z$, $K_{zz}$] space explored in this work. Overall, the trends with $T_{eq}$ are qualitatively similar across different compositions and mixing strengths. At the lower end of the $T_{eq}$ interval, O is mostly in H$_2$O and condensed magnesium silicates (showed as a single curve labelled as MgSiO$_x$) for all cases except the most sluggish one ($K_{zz}=10^6$ cm$^2$s$^{-1}$). At intermediate $T_{eq}$, CO rises, gas-phase SiO appears and disequilibrium models also show a localized island of Mg(OH)$_2$. Finally, at the upper end of the $T_{eq}$ interval, CO becomes the dominant carrier, SiO grows into the main Si- and O-bearing species, and MgSiO$_x$, mostly in the form of forsterite, is either absent (in disequilibrium) or declining (in equilibrium).

While the trend is similar in equilibrium and disequilibrium, there are some important differences. In equilibrium, [MgSiO$_x$]$_\text{O}$ remains mostly constant up to $T_{eq} \sim 1300$ K, with higher metallicities extending the magnesium silicates stability to higher temperatures. On the other hand, in disequilibrium, MgSiO$_x$ begins disappearing at $T_\text{eq} \sim 900$ K and vanishes completely for $T_\text{eq} \ge 1100$ K, regardless of metallicity. A higher $K_{zz}$ modestly increases [MgSiO$_x$]$_\text{O}$ in the $T_{eq}=900$ and 1000 K cases, without changing the overall shape of the curve. In the most sluggish mixing regime tested here ($K_{zz}=10^6$ cm$^2$s$^{-1}$), gravitational settling dominates over eddy diffusion throughout most of the column, causing silicates to sediment near the 1000 bar bottom (as illustrated in Fig.~\ref{fig:vertprof_condensates}), depleting the atmosphere of silicates and driving [MgSiO$_x$]$_\text{O}$ to negligible values even at $T_{eq}=700$ K.

The breakdown of refractories in disequilibrium directly shapes the non-monotonic behavior of H$_2$O. In the $900 \le T_{eq} < 1100$ K interval, minor species like Mg(OH)$_2$ and SiO scoop around half of the released O atoms and keep the amount of O bound to Mg and Si at $\sim 10\%$, while the remaining shifts into H$_2$O. At $T_{eq}=1100$ K, where also Mg(OH)$_2$ dissociates, [H$_2$O]$_\text{O}$ is pushed to 90-95\%. Consequently, while [CO]$_\text{O}$ increases monotonically with temperature, [H$_2$O]$_\text{O}$ first rises as silicates \textcolor{black}{become unstable and evaporate,} and then declines as the steady state shifts toward CO. Increasing $K_{zz}$ from $10^6$ to $10^{10}$~cm$^2$\,s$^{-1}$ shifts both the peak of [H$_2$O]$_\text{O}$ and the [H$_2$O]$_\text{O}$-[CO]$_\text{O}$ crossover point to higher temperatures by $200$~K. The equilibrium crossover lies $\sim 100$~K below the $K_{zz}=10^6$~cm$^2$\,s$^{-1}$ disequilibrium case, forming a monotonic sequence toward increasingly out-of-equilibrium regimes.

Finally, metallicity impacts equilibrium and disequilibrium cases differently. In equilibrium, higher metallicity enhances oxygen sequestration in refractories ([MgSiO$_x$]$_\text{O}$ rises from $15\%$ to $23\%$ at $700$~K, and from $1\%$ to $17\%$ at $1500$~K), suppressing [SiO]$_\text{O}$ at the high end of the $T_{eq}$ range, and [H$_2$O]$_\text{O}$ everywhere. In disequilibrium, composition has a much weaker effect: while [MgSiO$_x$]$_\text{O}$ increases at $700$~K from 13-15\% to 21-23\%, silicates consistently follow the same trajectory toward complete disappearance at $T_\text{eq} \sim 1100$~K.

Iron, in the form of Fe(OH)$_2$, was expected to be an oxygen carrier based on our previous equilibrium chemistry calculations in \citetalias{fonte23}. This is indeed the case when Fe is not allowed to condense. In such a scenario (not shown in the plots), Fe(OH)$_2$ captures between 8\% and 14\% of the O atoms at 700 K, depending on the metallicity. It decreases in concentration and then disappear at 1100-1200 K, at a $\sim 100$ K higher $T_{eq}$ than Mg(OH)$_2$. Instead, when condensation is turned on, Fe mostly disappears from the gas phase, thus ceasing to contribute significantly to the O partition, irrespectively from the other atmospheric parameters.

\subsection{Oxygen partition in the upper atmosphere}
\label{ss:results_upper}

The partition of oxygen in the upper atmosphere is similar to that in the lower atmosphere. In Fig.~\ref{fig:oxypart_upper}, we show the partition for the lowest metallicity case, as an example. The three-phases trend with $T_{eq}$ visible in the lower atmospheric region is reproduced in chemical equilibrium. Similarly, [MgSiO$_x$]$_\text{O}$ is identical in the two layers, since the silicates cloud deck in equilibrium extend up into the [10$^{-4}$, 10$^{-3}$] bar region. The behavior does not change across different compositions (not shown). The disequilibrium results also reproduce the trends found in the lower region. The most notable difference is that the upper region is cloud-free both for $K_{zz}=10^{6}$ and $10^{8}$ cm$^2$s$^{-1}$, since the mixing cannot sustain the 1 $\mu$m grains at sub-mbar pressures, as confirmed also by Fig.~\ref{fig:vertprof_condensates}. At the high temperature extreme of the $T_{eq}$ interval, equilibrium and disequilibrium results, irrespectively of the mixing strength, tend to converge to the same value, just as observed in the lower atmospheric region.

\begin{figure}
\centering
\includegraphics[width=0.95\columnwidth]{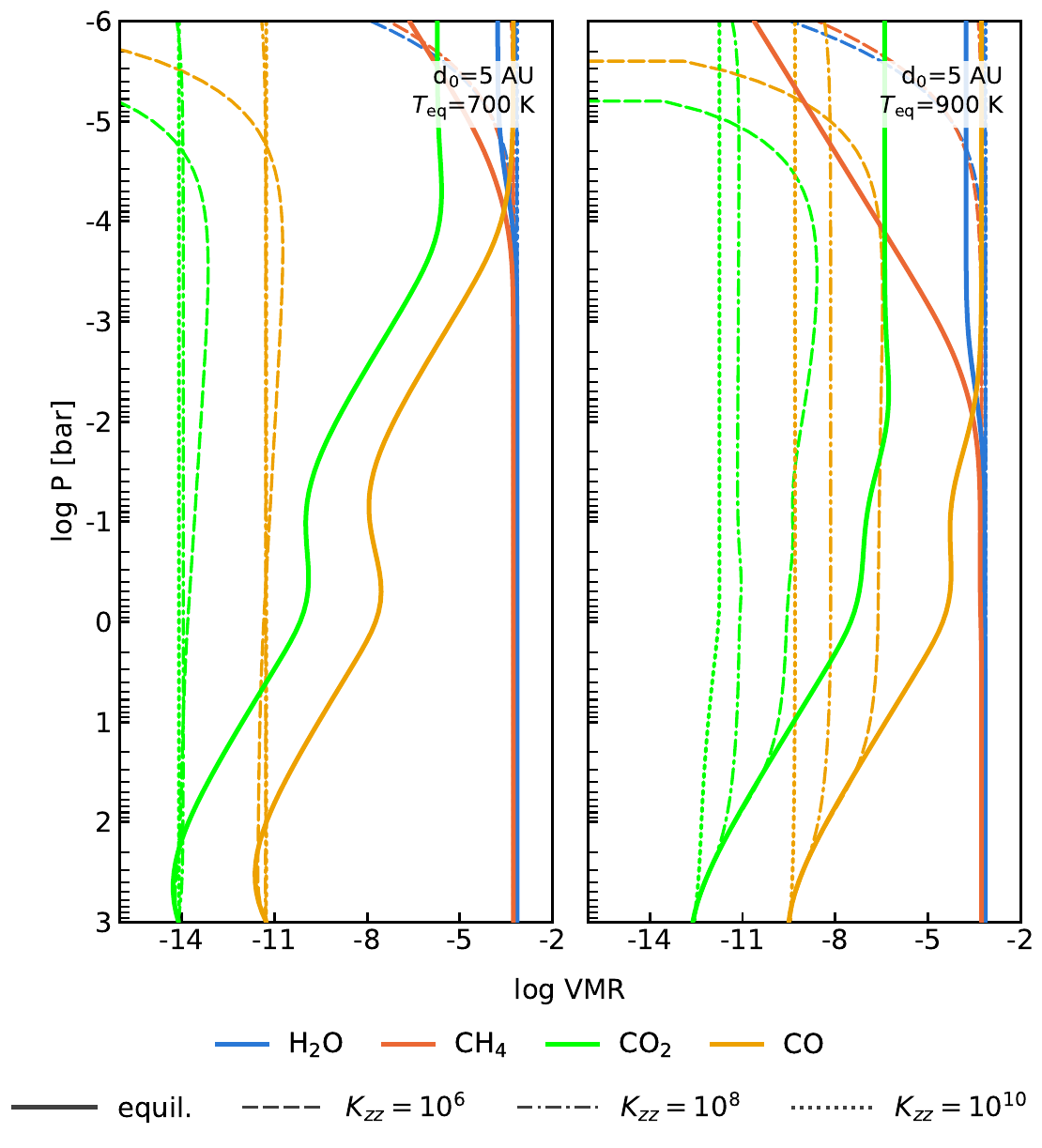}
\caption{The vertical VMR distribution of four observationally relevant gas phase species for the lowest metallicity case (i.e., the 5 AU scenario) at two different $T_{eq}$: 700 K on the left, 900 K on the right. Solid lines refer to the equilibrium chemistry results. Dashed, dash-dotted and dotted lines refer, respectively, to cases with $K_{zz}$ equal to 10$^6$, 10$^8$ and 10$^{10}$ cm$^2$s$^{-1}$.}
\label{fig:vert_profile_gases}
\end{figure}

If we focus on the two most abundant carriers, H$_2$O and CO, we see that the crossover point in temperature is shifted to lower temperature values for the equilibrium atmosphere with respect to the [0.1, 1] bar layer, but this does not happen in the disequilibrium one. The reason becomes clear by looking at Fig.~\ref{fig:vert_profile_gases}, which shows the vertical VMR profiles for H$_2$O, CO, CO$_2$ and CH$_4$. Vertical mixing erases the differences in concentration between the upper and the lower atmospheric region, at least in the two most abundant carriers (CO and H$_2$O) across all the temperatures and compositions tested here.

\subsection{Oxygen deficit and C/O bias}
\label{ss:oxygen_deficit}

\begin{figure*}
\centering
\begin{minipage}[c]{0.81\textwidth}
\centering
\includegraphics[width=0.48\textwidth]{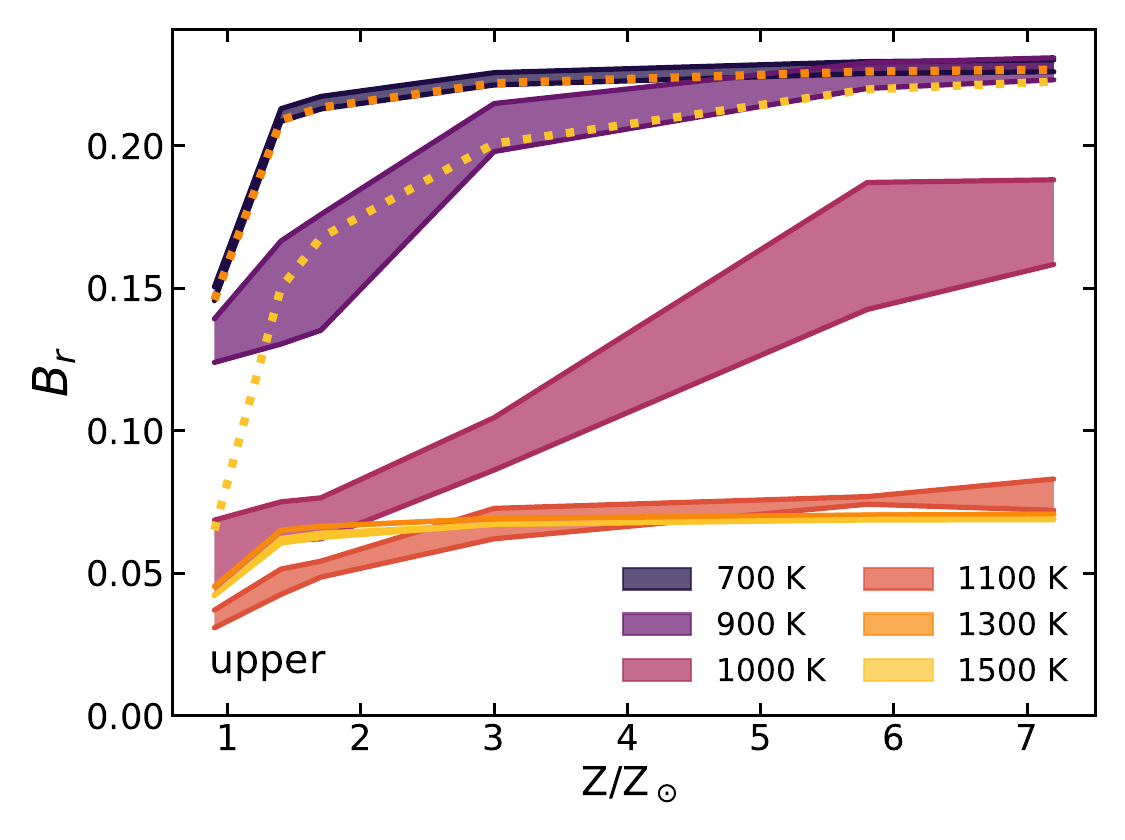}
\includegraphics[width=0.48\textwidth]{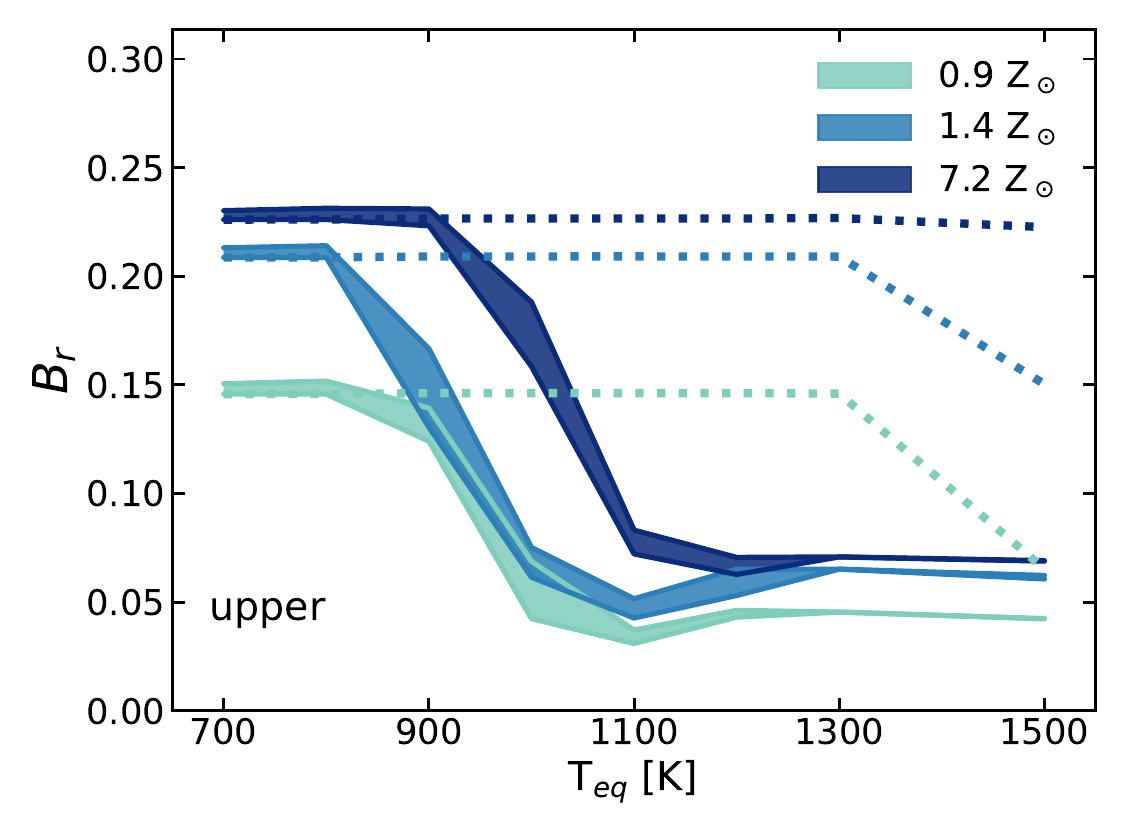}\\
\includegraphics[width=0.48\textwidth]{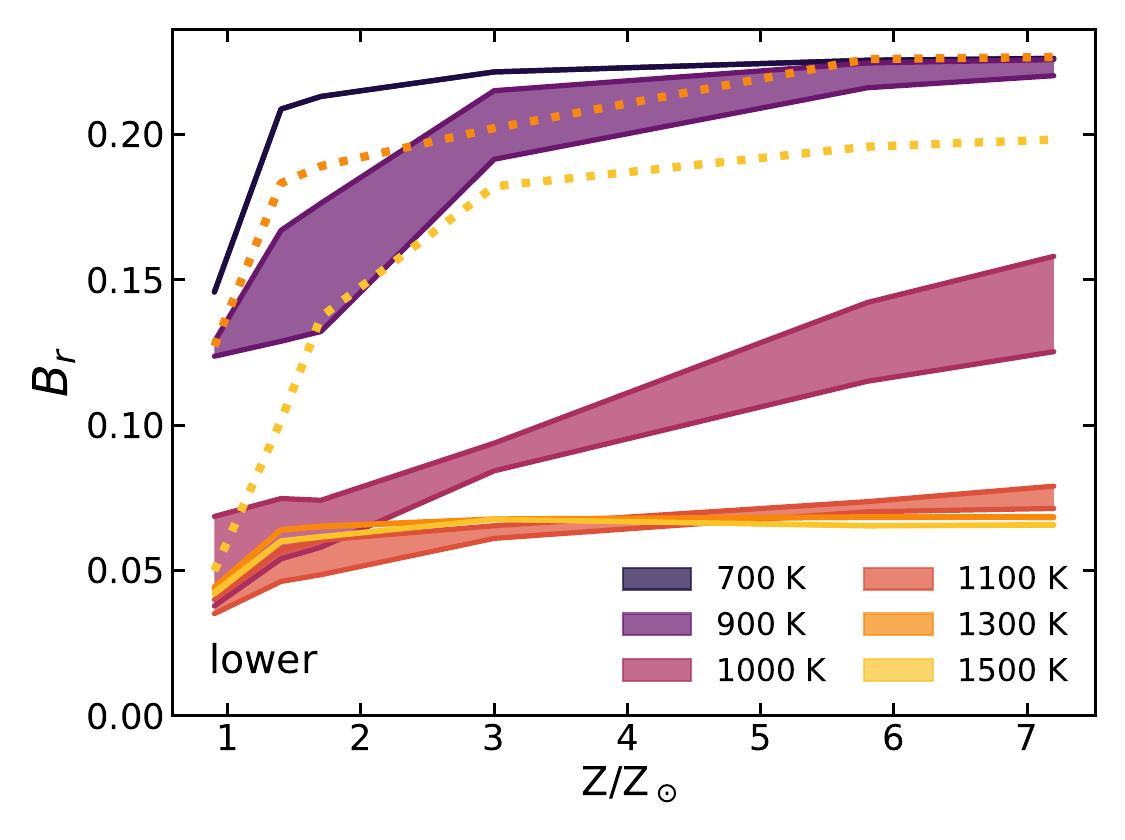}
\includegraphics[width=0.48\textwidth]{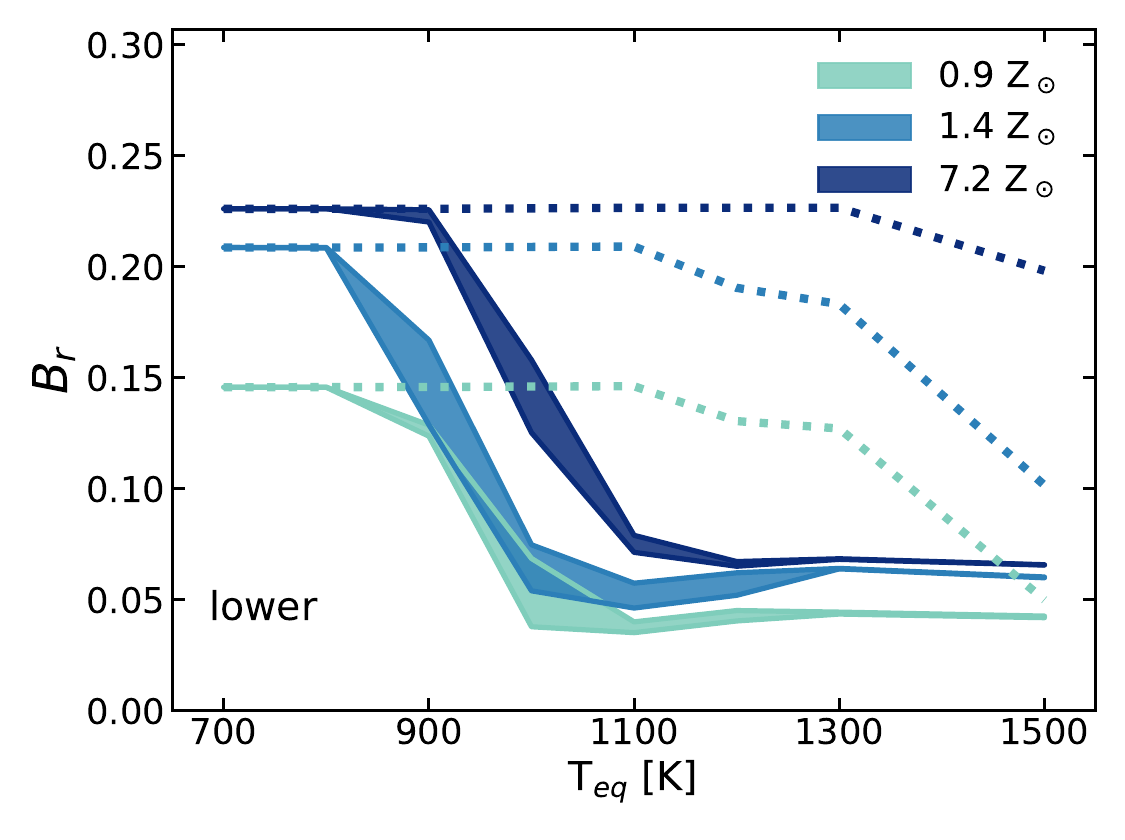}\\
\end{minipage}%
\hfill
\begin{minipage}[c]{0.189\textwidth}
\small
\caption{The C/O bias $B_r$ as a function of the planetary atmospheric metallicity (left) and the equilibrium temperature (right) in the upper (first row) and lower (second row) atmosphere for a selection of cases. Shaded areas represent the envelope of results for different $K_{zz}$. Dotted curves represent the equilibrium chemistry results, for the same color-coded case. Notice that in the panels on the left, only the equilibrium curves for $T_{eq}=1300$ and 1500 K are shown.\label{fig:oxydef}}
\end{minipage}
\end{figure*}

The C/O ratio derived by near infrared observations, $R_{\text{obs}}$, is quantified from the VMRs of species with sufficiently strong absorption bands in that region, namely CO, CO$_2$, CH$_4$ and H$_2$O. It is derived by the following formula:
\begin{equation}
R_{\text{obs}} = \frac{[\text{CO}]+[\text{CO}_2]+[\text{CH}_4]}{[\text{H}_2\text{O}]+[\text{CO}]+2[\text{CO}_2]}.
\end{equation}
where [X] represents the VMR of species X. However, as shown in \citetalias{fonte23}, refractory elements can sequester a substantial amount of O atoms, thus causing $R_{\text{obs}}$ to diverge from the \textcolor{black}{local} intrinsic C/O ratio of the exoplanetary atmosphere, $R$:
\begin{equation}
R = \frac{[\text{CO}]+[\text{CO}_2]+[\text{CH}_4]+[\text{MC}]}{[\text{H}_2\text{O}]+[\text{CO}]+2[\text{CO}_2]+[\text{MO}]},
\end{equation}
where [MC] and [MO] indicate the total number densities of C- and O-bearing species other than CO, CO$_2$, CH$_4$ and H$_2$O. In principle, also carbon can be sequestered by refractory elements or in refractory phases, such as graphite \citep[e.g.][]{woitke18}, but equilibrium runs in this work and in \citetalias{fonte23} confirm that this is not the case for the C/O ratio values explored here.
If $[\text{MO}] > 0$, then $R_{\text{obs}} > R$, and the divergence between the two can be quantified by the oxygen deficit $\delta_O$:
\begin{equation}
\delta_O = 1 - \frac{R}{R_{\text{obs}}}.
\end{equation}
as defined in \citetalias{fonte23}. By substituting the $R_{\text{obs}}$ and $R$ definitions in the expression for $\delta_O$, it is possible to see that $\delta_O = [\text{MO}]_\text{O}$, where $[\text{MO}]_\text{O} = [\text{MO}]/([\text{H$_2$O}] + [\text{CO}] + 2[\text{CO}_2] + [\text{MO}])$. If there is essentially no free oxygen, as it is the case in our calculations, then the dividend in the $[\text{MO}]_\text{O}$ expression represents the entire reservoir of O, thus it can be rewritten as $[\text{MO}]_O = [\text{MO}]/[\text{O}]$. In all the cases explored here, $[\text{MO}]_\text{O}$ is equal to the fraction of O atoms sequestered by Mg and Si.

The above oxygen deficit, however, does not account for vertical transport. In the disequilibrium runs sedimentation can move oxygen carriers from one layer to another, thus skewing the outcome when it is expressed in terms of $\delta_O$. As such, we introduce another quantity, called C/O bias, $B_r$. This quantity compares $R_{\text{obs}}$ as calculated by the results of a VULCAN run using the full \texttt{SILANE} network, \textcolor{black}{namely $R_{\text{obs}}^{full}$}, and $R_{\text{obs}}$ as calculated by the results of a FastChem run with the minimal set of molecules, i.e. H$_2$O, CO, CO$_2$ and CH$_4$, \textcolor{black}{namely $R_{\text{obs}}^{ideal}$}. The C/O ratio is calculated only from H$_2$O, CO, CO$_2$ and CH$_4$ in both cases. However, the VULCAN run includes all the C and O carriers, and both C and O can be vertically transported. The FastChem run, instead, considers the four visible carriers as the only carriers of C and O. In analogy with $\delta_O$, $B_r$ it is defined as
\begin{equation}
B_r = 1 - \frac{R_{\text{obs}}^{ideal}}{R_{\text{obs}}^{full}}
\end{equation}
and estimates how much skewed is the observed C/O ratio with respect to the intrinsic one. If no C and O exists in atomic form, as it is the case in the $T_{eq}$ interval studied in this work, then $R_{\text{obs}}^{ideal}$ is equal to the the intrinsic C/O ratio set by Table $\ref{tab:element_setting}$ abundances. Notice that, despite the fact that $B_r$ is sensitive to the global state of the atmospheric column, it is still a local quantity, since it stems for locally evaluated VMRs values.

Our results in terms of $B_r$ are shown in Fig.~\ref{fig:oxydef}. The first feature we notice is that the trend with metallicity has a single, sharp increase in $B_r$ when moving from the 0.9 $Z_\odot$ scenario to the 1.4 $Z_\odot$ scenario. This is explained by the two scenarios having the strongest increase in the refractory content (+125\%) with respect to that in oxygen abundance (+56\%). The impact of the enlarged stock of Mg and Si is mostly felt at the lower end of the explored $T_{eq}$ range, where $B_r$ increases from $\sim$0.145 to $\sim$0.23 at 700 K irrespectively of the vertical mixing intensity. In contrast, further increases in metallicity have the strongest impact at intermediate ($\sim 1000$ K) temperatures, where $B_r$ gradually rises from 0.05-0.08 at $Z=1.4$ $Z_\odot$ to 0.12-0.19 at $Z=7.2$ $Z_\odot$. This trend holds for both the upper and the lower atmosphere. At the upper end of the temperature interval, i.e. for $T_{eq} \ge 1300$ K, $B_r$ ranges between 0.045 and 0.07.

The second feature is that most of the variability induced by vertical dynamics manifests itself at intermediate temperatures. Outside the $800 < T_{eq} \le 1200$ K interval, $B_r$ is essentially unsensitive to $K_{zz}$, especially in the lower atmospheric region. This is clearly visible in the right-hand panels of Fig.~\ref{fig:oxydef}, where the dependence of $B_r$ on $T_{eq}$ is shown for three compositions: the lowest and highest metallicity ones, and the metallicity that corresponds to the largest $B_r$ jump discussed in the previous paragraph ($Z=1.4$ $Z_\odot$). The spread induced by vertical mixing at intermediate $T_{eq}$ causes $B_r$ to become partially degenerate in $Z$, with even the cases with 0.9 and 1.4 $Z_\odot$ showing a partial overlap.

Equilibrium chemistry results are shown in Fig.~\ref{fig:oxydef} as dotted curves. The left-hand panels show only the curves for $T_{eq} = 1300$ and 1500 K, since all the others would overlap on the black-shaded area. As we have already discussed in Sect.~\ref{ss:results_lower}, magnesium silicates persist up to substantially higher $T_{eq}$, since there is no vertical settling, which keeps $B_r$ constant and at the highest value within each composition. There is a strong dependence on metallicity of the high-temperature $B_r$ values in equilibrium: at the lower end of the $Z$ interval, equilibrium and disequilibrium calculations both find low (0.04 to 0.06) $B_r$, while at high $Z$ the $B_r$ is essentially constant in temperature at 0.23. The right-hand panels of Fig.~\ref{fig:oxydef} underline again this feature, by showing a constant $B_r$ up to 1200 K.

\begin{figure*}
\centering
\begin{minipage}[c]{0.81\textwidth}
\centering
\includegraphics[width=0.99\textwidth]{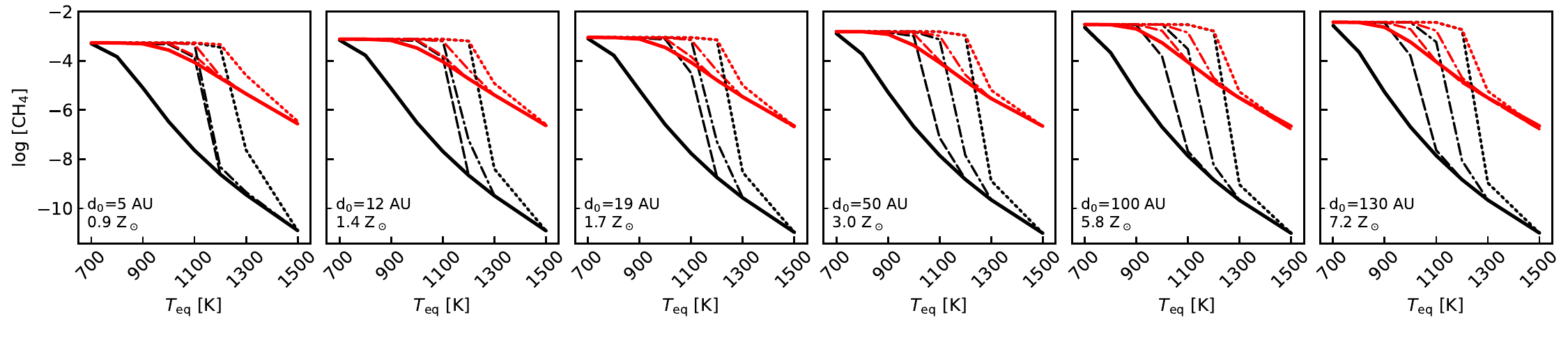}\\
\includegraphics[width=0.99\textwidth]{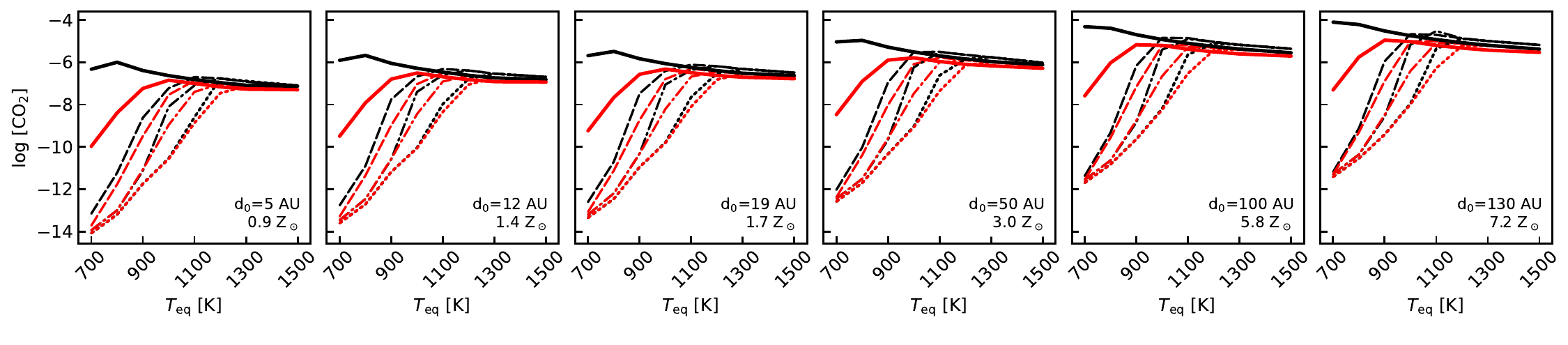}\\
\includegraphics[width=0.99\textwidth]{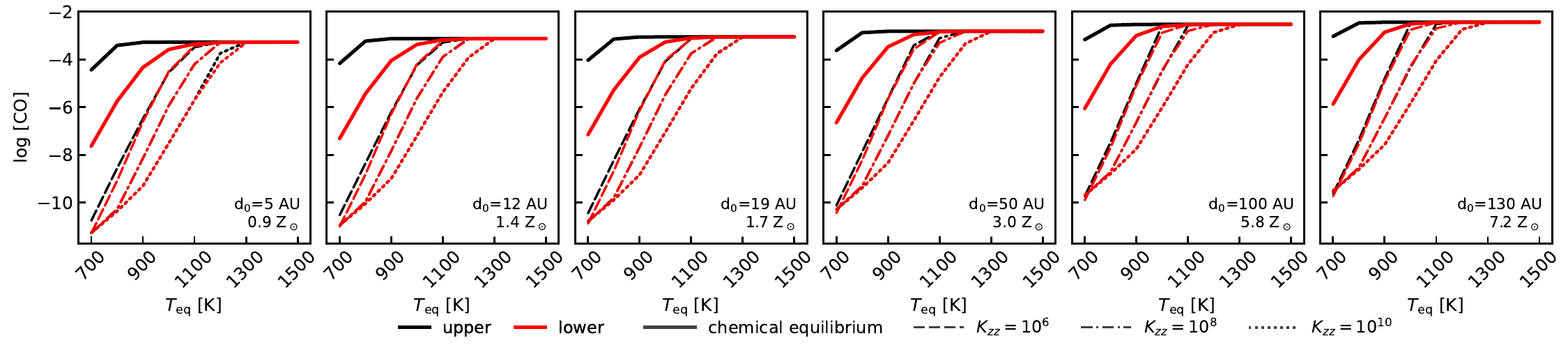}\\
\end{minipage}%
\hfill
\begin{minipage}[c]{0.189\textwidth}
\small
\caption{The mixing ratios of CH$_4$ (first row), CO$_2$ (second row) and CO (third row) as a function of the equilibrium temperature for the six formation scenarios explored in this work. Black curves refer to the upper atmospheric region and red curves to the lower atmospheric region. Dashed, dash-dotted and dotted lines refer to different vertical mixing intensities, with $K_{zz}$ equal to 10$^6$, 10$^8$ and 10$^{10}$ cm$^2$s$^{-1}$ respectively. Solid lines refer to equilibrium chemistry results.\label{fig:minor}}
\end{minipage}
\end{figure*}

\subsection{Other species of interest}
\label{ss:other_species}

We explore the impact of our new chemical network on the mixing ratios of two additional molecules that participate in the C/O ratio calculation, CH$_4$ and CO$_2$. These molecules are observationally relevant because of their relatively high concentration and their prominent absorption features in the near- and mid-infrared. Their mixing ratios in the upper and lower atmospheric regions explored in this work are showed in Fig.~\ref{fig:minor}.

The CH$_4$ VMR is mostly constant with respect to temperature for $T_{eq} \le 1000$ K in the lower atmospheric region. In disequilibrium, this is true also for the upper region and for all the tested $K_{zz}$. In that temperature range, CH$_4$ is the main C-bearing species, carrying nearly 100\% of the C atoms. Above 1000 K, differences start to arise, with well mixed atmospheres showing a higher amount of CH$_4$ up to 1200 K. The upper atmospheric region (black) rapidly depletes of CH$_4$ when $K_{zz} \le 10^{8}$ cm$^2$s$^{-1}$, while the lower atmospheric region (red) shows a gentler decrease. In Jupiter-like atmospheres, CH$_4$ is favored over other C-carriers like CO and CO$_2$ both at low temperatures and high pressures. Under our specific choice of the pressure-temperature profile (see Sect.~\ref{ss:atmo_profiles}), the relatively shallow temperature gradient in the atmosphere favors the production of CH$_4$, which explains why disequilibrium runs are far more CH$_4$-rich than equilibrium ones.

The CO$_2$ VMR, instead, is mostly constant for $T_{eq} \ge 1200$ K, decreasing by only 0.2-0.3 dex at 1500 K. This behaviour is found at both atmospheric layers and for both equilibrium and disequilibrium. On the other hand, for $T_{eq}$ in [700, 1200], its concentration depends decisively on the mixing strength. Out-of-equilibrium atmospheres are depleted of CO$_2$, with differences between the equilibrium case and the $K_{zz}=10^6$ cm$^2$s$^{-1}$ case that can be as large as $\sim6.5$ dex in the upper atmosphere. The disequilibrium concentrations at $T_{eq}=700$ K are degenerate in $K_{zz}$, but for $800 \le T_{eq} < 1200$ K they are separated by 1-1.5 dex, which gives the opportunity to establish an upper limit on the mixing strength in the case of a non-detection.

Both CH$_4$ and CO$_2$ have VMRs that can be up to two orders of magnitude apart in disequilibrium models when measured in the two atmospheric regions considered here. However, in Sect.~\ref{ss:results_upper} we remarked that the oxygen partition does not seem to differ substantially, apart for the disappearance of the magnesium silicates in the intermediate ($K_{zz} = 10^{8}$ cm$^2$s$^{-1}$) mixing case. So, to be sure that this is not an artefact due to the concurrent drift of more than one species, in the lower row of Fig.~\ref{fig:minor} we also show the VMR of CO. At variance with the other two molecules, it does, in fact, have the same concentration at both atmospheric levels for most of the [$T_{eq}$, $Z$, $K_{zz}$] parameter space tested. The maximum difference occurs at $T_{eq} = 700$ K and for the $K_{zz} = 10^{6}$ cm$^2$s$^{-1}$ case, where CO is up to 0.52 dex more abundant in the low-metallicity scenarios.

\begin{figure}
\centering
\includegraphics[width=0.90\columnwidth]{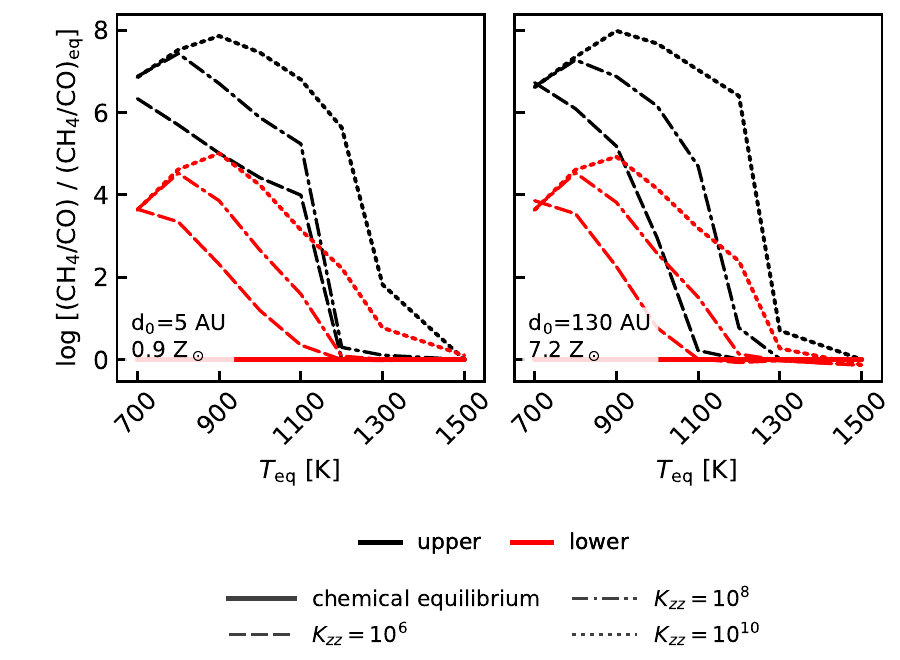}
\caption{The ratio between the concentrations of CH$_4$ and CO, normalized to their value at equilibrium for two different compositions: 0.9 $Z_\odot$ on the left, 7.2 $Z_\odot$ on the right. Black curves refer to the upper atmospheric region and red curves to the lower atmospheric region. Dashed, dash-dotted and dotted lines refer to different vertical mixing intensities, with $K_{zz}$ equal to 10$^6$, 10$^8$ and 10$^{10}$ cm$^2$s$^{-1}$ respectively.}
\label{fig:minor_ratio}
\end{figure}

The different sensitivity of CH$_4$, CO$_2$ and CO to the vertical mixing intensity makes it possible to use their ratios to constrain $K_{zz}$. As an example, in Fig.~\ref{fig:minor_ratio} we show how the disequilibrium [CH$_4$]/[CO] ratio, normalized to same ratio evaluated in equilibrium, changes with $T_{eq}$. Apart for the high temperature extreme, it is always possible to discern the disequilibrium scenario from the equilibrium one, and if $800 \le T_{eq} < 1200$ K, it is also possible to estimate $K_{zz}$. The difference is maximized precisely in the $T_{eq}$ interval where $B_r$ shows the maximum sensitivity to vertical transport. This molecular ratio differs when measured in the upper and the lower atmospheric layer, potentially allowing for two independent measurements to constrain the same quantity. Finally, by comparing the left- and right-hand panels of Fig.~\ref{fig:minor_ratio}, we notice that [CH$_4$]/[CO] is only marginally sensitive to the intrinsic metallicity, at least in the range tested here.

\section{Discussion}
\label{sec:discussion}

\subsection{The impact of metallicity on the vertical thermal profile}
\label{ss:metopac}

\begin{figure}
\centering
\includegraphics[width=0.95\columnwidth]{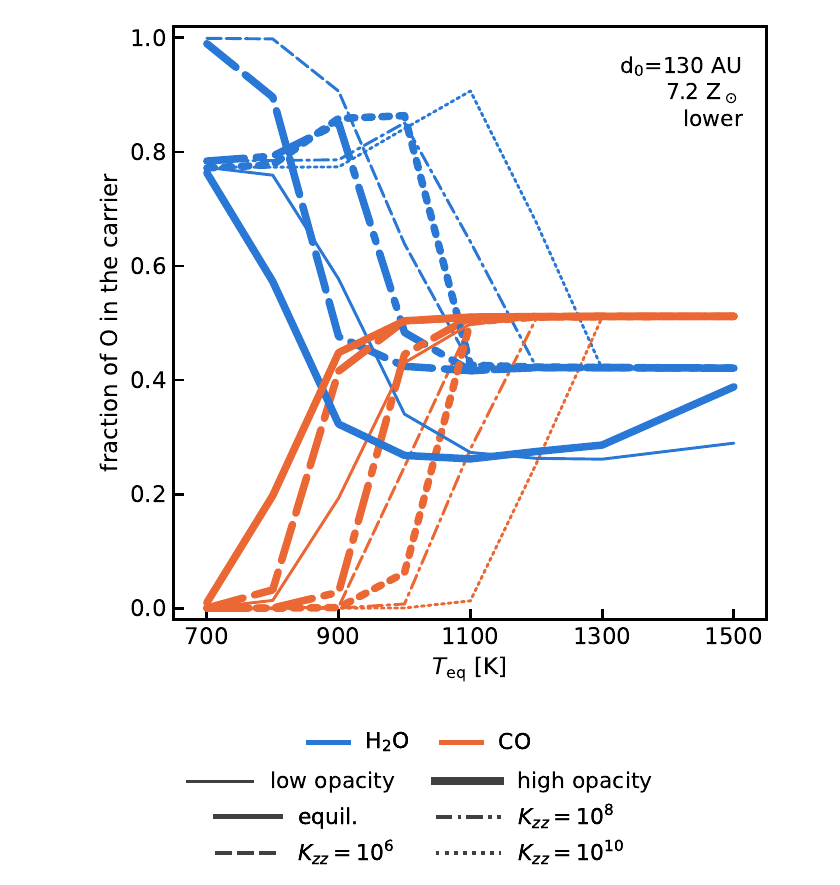}
\includegraphics[width=0.82\columnwidth]{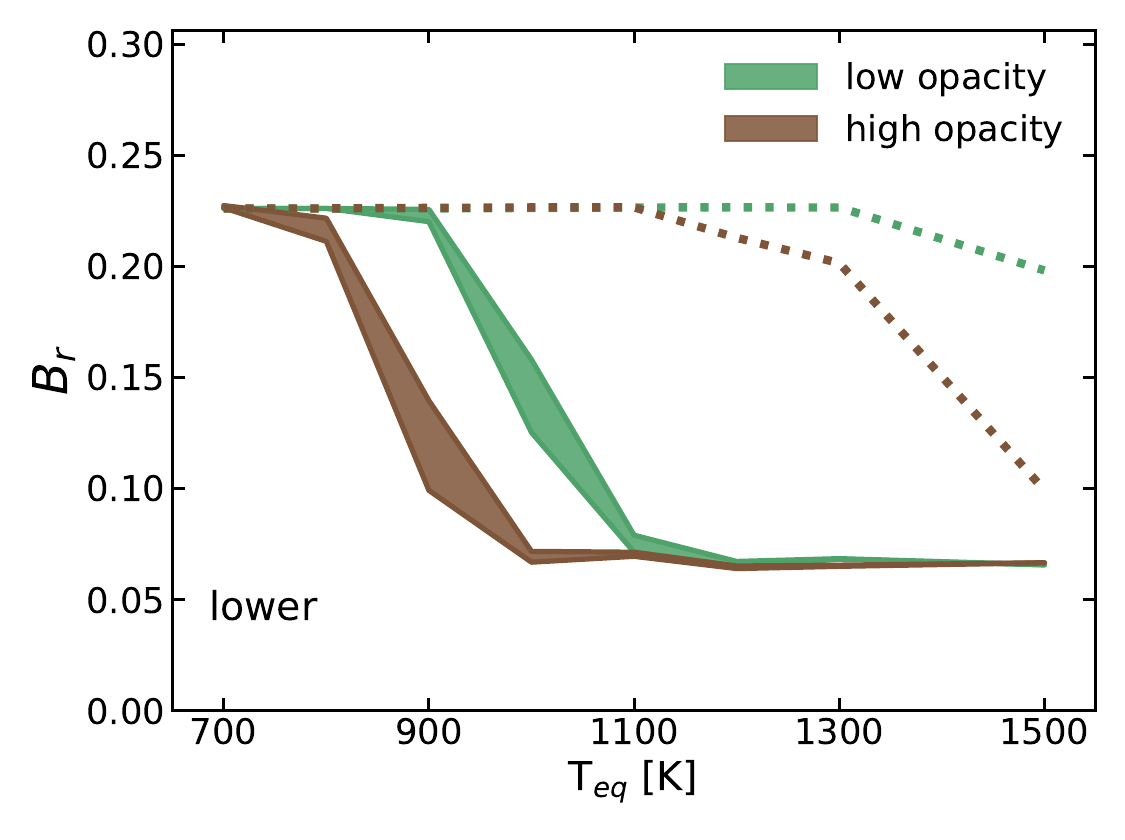}\\
\caption{The impact of changing the vertical temperature profile on the oxygen partition in the lower atmospheric region. Upper: the fraction of oxygen locked in its two main carriers. Thin lines refer to the cases that used the set vertical profiles shown in Fig.~\ref{fig:temp_profiles}, while thick lines refer to the cases for which we doubled the infrared opacity of the atmosphere, as described in Sect.~\ref{ss:atmo_profiles}. Lower: the C/O bias $B_r$ for the $Z=7.2$ $Z_\odot$ case using the same infrared opacity as the rest of the paper (green) and using a doubled infrared opacity (brown). Dotted lines refer to equilibrium models with the same atmospheric profile.}
\label{fig:oxydef_highopac}
\end{figure}

As introduced in Sect.~\ref{ss:atmo_profiles}, increasing the metallicity increases the concentration of infrared absorbers within the atmospheric column, which in turn changes the vertical thermal profile. We did not couple VULCAN to a radiative transfer code (such as HELIOS), since this analysis was outside the scope of this paper. Thus, strictly speaking, the vertical thermal profiles that we adopted are not self-consistent, since we kept the $\kappa_{th}$ term in Eq.~\ref{eq:guillot} constant even if we varied the elemental makeup of the atmosphere. To gain insight on the impact of this parameter, we tested a set of profiles obtained by doubling the infrared opacity.
The lower atmospheric region is the most impacted by the change in opacity, as shown in Fig.~\ref{fig:oxydef_highopac} where we report the oxygen partition and the oxygen deficit.

Increasing the opacity heats up the atmosphere at all layers. As such, the formation of CO is favored and its mixing ratio starts to increase at lower $T_{eq}$. CO becomes the main O carrier at $T_{eq}$ between 100 and 150 K lower than in the baseline opacity scenario (left panel of Fig.~\ref{fig:oxydef_highopac}). This change is amplified in presence of stronger mixing. The impact on $B_r$ follows the same pattern: since the temperature is higher, the survival of magnesium silicates (and of other Mg and Si molecules) is disfavored. As such, at any given $T_{eq}$, $B_r$ is lower than in the reference opacity case and the impact is amplified by a stronger mixing, while the overall trend similar.

\subsection{The role of Fe}
\label{ss:iron}

 In our model iron is nearly entirely locked up in metallic form at both atmospheric levels and thus it does not contribute to the oxygen deficit. As mentioned in Sect.~\ref{ss:results_lower}, this result reproduces the behaviour found by \citet{visscher10} at 1 bar and \citet{fegley10} at 10$^{-4}$ bars by using chemical equilibrium models with condensation. On the other hand, it differs substantially from the \citetalias{fonte23} results, since their model did not include condensation and thus found that most Fe forms Fe(OH)$_2$, at least for $T_{eq}\lesssim1100$ K. Recently, \citet{kiefer24} used a detailed kinetic cloud formation model coupled with an expanded gas phase chemical kinetics network to investigate the nucleation and bulk growth of mineral grains in an hot Jupiter atmosphere. In their model, most Fe ended up in the grains as either fayalite (the Fe endmember of the olivine series, Fe$_2$SiO$_4$) or as Fe$_2$O$_3$, depending on the temperature of the atmospheric layer. Their conclusions are supported by the fact that the Fe supersaturation required to condense the metallic iron is very high and thus, despite being the thermodynamically favored product in this temperature range, it might not form \citep{tielens22}. However \citet{kiefer24} did not include the vertical mixing, since they studied one (p,T) point within the atmosphere at time and, more importantly, they did not include metallic iron formation. The different goals and assumptions of these studies make it difficult to compare their results, even though they suggest that Fe might have a role in O sequestration in scenarios where atmospheric general circulation favors transient states over the equilibrium end-state.

\section{Conclusions}
\label{sec:conclusions}

In this paper, we used our new Mg-Si-Fe chemical network, \texttt{SILANE} \citep{simonetti26}, coupled with the S-N-C-H-O VULCAN network, to explore how oxygen partitions among its main carriers in two observationally important atmospheric layers, [0.1, 1] bar and [10$^{-4}$, 10$^{-3}$] bars. The \texttt{SILANE} network includes the condensation of the two most abundant magnesium silicates expected from chemical equilibrium calculations, enstatite and forsterite, plus metallic iron. We carried out our calculations starting from a set of non-Solar, self-consistently derived initial atmospheric compositions. Our main goal was to quantify how oxygen sequestration in non-observed species skews the C/O ratio $R$ when derived from H$_2$O, CO, CO$_2$ and CH$_4$ alone. We quantified this deviation using the C/O bias $B_r$. While $B_r$ depends on both O and C sequestration, we showed that it mostly depends on O sequestration for the compositions tested here. We then explored the effect of changing the planetary equilibrium temperature in the [700, 1500] K range and the vertical eddy mixing coefficient in the [10$^{6}$, 10$^{10}$] cm$^2$s$^{-1}$. Finally, we also investigated the mixing ratios of two observationally relevant species, CO$_2$ and CH$_4$, which can be used to constrain the vertical mixing strength.

Our main conclusions are as follows:
\begin{enumerate}
\item For the set of compositions tested here, $B_r$ is comprised between 0.145-0.23 at $T_{eq} = 700$ K, and 0.045-0.07 at $T_{eq} = 1500$ K. $B_r$ is sensitive to the vertical atmospheric dynamics in the $800 < T_{eq} \le 1200$ K interval. This implies that warm ($T_{eq}\le800$ K) exoplanets systematically have a lower intrinsic $R$ than observed, transitional hot ($800 < T_{eq} \le 1200$ K) exoplanets might or might not have an lower $R$ depending on the atmospheric dynamics, and hot ($T_{eq}>1200$ K) exoplanets have only a minimally skewed $R$.
\item There is a common trend in the oxygen partition, both in equilibrium and disequilibrium models. However, while $B_r$ is constant between 700 K to 1100-1300 K in equilibrium models, disequilibrium models show a rapid drop starting already at 900 K.
\item Since the dissolution of MgSiOx starts at a lower $T_{eq}$ in the disequilibrium models, Mg(OH)$_2$ has the opportunity to form. This is never the case in the equilibrium models when magnesium silicates are allowed to form, although this species can form if only gas-phase refractory carriers are allowed, as in \citetalias{fonte23}.
\item Higher metallicities produce higher $B_r$ across the entire $T_{eq}$ range, so that the net effect is a shift with no structural changes.
\item $B_r$ is mostly independent of the probed atmospheric layer. Only a subset of simulations at $T_{eq} = 900$ and 1000 K show non-negligible ($\lesssim 0.04$) differences.
\item The dependence of $B_r$ on the exact atmospheric structure, here studied by doubling the mean infrared opacity of the atmosphere, is in the form of a shift of the trend towards lower $T_{eq}$ by 100-150 K.
\item When allowed to condense, Fe is not an O carrier for any of the combinations tested in this work. If Fe condensation is switched off, Fe forms Fe(OH)$_2$ for $T_{eq} \lesssim 1100$ K, which can carry $\sim8-14\%$ of the O stock, increasing $B_r$ by a factor of $\sim 1.5$.
\end{enumerate}

These results clearly show that the best targets for elemental abundance characterization are those with either $T_{eq} \lesssim 800$ K, or $T_{eq} \gtrsim 1300$ K since, in these cases, vertical mixing can be ignored and $B_r$ can be estimated either by equilibrium chemistry (for the low $T_{eq}$ end), or by considering only SiO as non-observed O carrier.

As already noted in Sect.~\ref{ss:oxygen_deficit}, these results are valid for relatively O-rich atmospheres. In C-rich atmospheres we expect significant C sequestration and substantially different trends. This exploration will be the goal of a future work.

\section*{Acknowledgements}

The Authors want to thank J.~Lyons, D.~Fainozzi and I. Crossfield for the insightful discussions in different moments of the drafting phase of this work. P.S. and D.P. are supported by the Italian National Institute of Oceanography and Experimental Geophysics (OGS) and CINECA through the HPC-TRES program awards N.~2022-02 and 2022-05 and by the National Research Centre for High Performance Computing, Big Data e Quantum Computing CN\_00000013, Spoke 3 `Astrophysics and Cosmos Observations' (CUP: C53C22000350006). The Authors acknowledge the support of the Italian National Institute of Astrophysics (INAF) through the Main Stream project `Ariel and the astrochemical link between circustellar discs and planets' (CUP: C54I9000700005) and the PRIN INAF 2019 `Platea' and `Hot-Atmos', of the Italian Space Agency (ASI) through the ASI-INAF contracts N.~2016-23-H.0 plus addendum 2016-23-H.2-2021 and 2021-5-HH.0 plus addenda 2021-5-HH.1-2022 and 2021-5-HH.2-2024 and of the European Research Council via the Horizon 2020 Framework Programme ERC Synergy `ECOGAL' Project GA-855130. The Authors also acknowledge the support of the European Cooperation in Science and Technology (COST) Action CA22133 `PLANETS'.


\bibliographystyle{mnras}
\bibliography{example}

\end{document}